\documentclass{acmsiggraph}
\usepackage[]{algorithm2e}
\usepackage{epstopdf}
\usepackage{color}

\definecolor{red}{rgb}{0.8,0,0}
\definecolor{darkred}{rgb}{0.6,0,0}
\definecolor{green}{rgb}{0.0,0.5,0}
\definecolor{blue}{rgb}{0,0,0.75}

\newcommand{\timeJitter}{t_{j}}
\newcommand{\FWHM}{H_m}
\newcommand{\DCR}{R_{dc}}
\newcommand{\sDeadTime}{t_o}

\DeclareMathOperator{\Poisson}{Pois}

\title{A Computational Model of a Single-Photon Avalanche Diode Sensor \\for Transient Imaging}

\author{Quercus Hernandez \qquad Diego Gutierrez  \qquad Adrian Jarabo \\ \\ Universidad de Zaragoza, I3A }
\pdfauthor{Hernandez et al.}

\TOGonlineid{45678}


\CopyrightYear{2017}
\setcopyright{acmcopyright}
\conferenceinfo{CONFERENCE PROGRAM NAME}{MONTH, DAY, and YEAR} 
\isbn{THIS-IS-A-SAMPLE-ISBN}\acmPrice{\$15.00}
\doi{http://doi.acm.org/THIS/IS/A/SAMPLE/DOI}

\begin{document}



\maketitle

\begin{abstract}

Single-Photon Avalanche Diodes (SPAD) are affordable photodetectors, capable to collect extremely fast low-energy events, due to their single-photon sensibility. This makes them very suitable for time-of-flight-based range imaging systems, allowing to reduce costs and power requirements, without sacrifizing much temporal resolution. In this work we describe a computational model to simulate the behaviour of SPAD sensors, aiming to provide a realistic camera model for time-resolved light transport simulation, with applications on prototyping new reconstructions techniques based on SPAD time-of-flight data. Our model accounts for the major effects of the sensor on the incoming signal. We compare our model against real-world measurements, and apply it to a variety of scenarios, including complex multiply-scattered light transport. 

\end{abstract}


%
%


\keywordlist

\conceptlist


\section{Introduction}
Transient imaging has recently emerged, enabling a wide range of applications for computer vision and scene understanding~\cite{Jarabo2017}.  Disambiguating light transport in the temporal domain has allowed capturing light in motion~\cite{Velten2013}, non-line-of-sight imaging~\cite{Velten2012} or reflectance acquisition~\cite{Naik2011}. However, the used imaging technology is in general too expensive, difficult to operate and time-consuming to be used in the wild. 

Photon counting technology, such as single-photon avalanche diodes (SPAD), is a promising technology to address some of these limitations. These detectors are able to detect ultrafast signals, in the order of picosecond resolution, with very high sensitivity, by producing an avalanche current reaction when activated by a photon. They have been demonstrated useful in several fields such as basic quantum mechanics \cite{Rarity1990,Shih1988}, measurements of fluorescent decays and luminescence in physics, chemistry, biology and material science \cite{Li1993,Soper1993}, or single molecule detection \cite{Matem1983,Andreoni1984}. 
More recently, they have been applied in the particular context of transient imaging, to determine photons time of flight in actively illuminated setups~\cite{Gariepy2015}. This has allowed to capture non-line-of-sight objects~\cite{Buttafava2015} with significantly cheaper capturing systems than the previous work~\cite{Velten2012}, reducing both the intensity of the scene illumination and the capture times, and in general the complexity and sensitivity of the system. 

\begin{figure}[ht]
  \centering
  \includegraphics[width=3.0in]{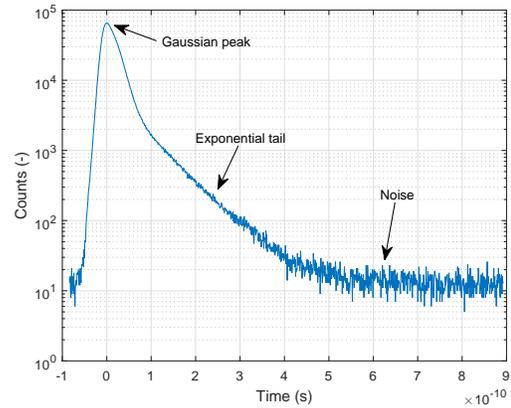}
  \caption{Measured temporal impulse response of a 20 $\mu$m CMOS SPAD with excess voltage of $V_{e}=20$ V. The curve shows two of the main characteristics of SPAD sensors, including the Gaussian plus exponential shape of the time jitter, as well as the effect of internal and background noise. A filtered normalized version of the curve is used as a data-driven pdf for modelling time jitter in our probabilistic model. }
  \label{fig:jitter}
\end{figure} 

However, SPADs have a number of negative effects which lead to signal deterioration, such as the time jitter or afterpulsing. Figure~\ref{fig:jitter} shows an example of the temporal response of a SPAD for an impulse signal. In this work we develop a computational model for SPADs, including all relevant effects on the time-resolved signal incoming the sensor, aiming the provide a predictive physically-plausible sensor model for transient light transport simulations. This is very important  for developing new SPAD-based reconstruction methods, as well as prototyping and assessing the limits of these reconstruction methods before building actual real-world tests. Our model accounts for effects such as the detection efficiency, time jitter, sensor's quenching, internal thermal noise, afterpulsing and pixel's crosstalk. We demonstrate the accuracy of our model comparing against real-world captured data, and apply it on top of complex light transport simulations~\cite{Jarabo2014} showing multiple diffuse interreflections.

\section{Related Work}

Here we focus on works computationally modeling the behavior of SPADs. We refer to other sources for details on SPADs~\cite{Charbon2007,Kirmani2014} and transient imaging in general~\cite{Jarabo2017}. 
The first work focusing on the simulation of SPAD detectors~\cite{Zappa2000} models the electronics of an active quenched circuit. Later, Dalla et al. \shortcite{Dalla2007} describes a circuit model with passive quenching which can be implemented in computer aided software. It can predict precisely the electronic behavior of the avalanche current ignition, quenching and recovery process, although it is restricted to passive quenching. Years later, Mita et al.~\shortcite{Mita2008} and Zappa et al.~\shortcite{Zappa2009} presented more complete electronic models including both active and passive quenching. However, these models focus on predicting the low-level electronics of the sensors, neglecting the counting process and the photon source so they cannot be used as camera models for simulation.

Repich et al.~\shortcite{Repich2009} implement a SPAD model for simulating time-resolved fluorescence decay measurements in a fluorophore solution using Monte Carlo simulation. The target of the paper is to show how imperfections of the sensor affect the results of the measurements. They model the main characteristics of the SPAD detector such as afterpulse, time jitter, dark counts and dead time. However, the dead time is naively implemented as a post-process of the total counts, which might result in rejecting a pulse but not its correlated afterpulse.

Guilinatti et al. \shortcite{Gulinatti2011} present a physically-based model of SPAD detectors that can accurately reproduce its temporal behavior. Both the photon detector probability and time jitter are fully modeled, but other effects such as afterpulse or quenching are not taken into account. The main goal of the paper is to understand the limitations of the current sensors and discuss some device modifications in order to overcome them.

Finally, Kazma et al.~\cite{Kazma2015} develop a temporal model of the photon arrival time in a SPAD detector. The light source is modelled as a monochromatic laser pulse following a Poissonian distribution. The dead time is applied to the sum of dark counts and source photons, followed by the afterpulsing probability as a post-process. The main drawback of Kazma's method is that it does not take into account the hold-off time of the avalanches generated by the afterpulses. Moreover, it does not model the time jitter of the sensor due to the low temporal resolution of the simulations (milliseconds).

As opposed to these works, our model predicts all temporal and spatial effects of the SPAD on the incoming light signal, including detection efficiency, time jitter, quenching, internal thermal and external noise, afterpulsing and pixel's crosstalk.

\section{A Computational SPAD Model}

A SPAD detector consists of a reversed biased p-n junction (diode) above its breakdown voltage ($V_{B}$), generating a high electric field. The sensor is in a semi-stable state, in which a single photon is able to start an avalanche through the electrons inside the semiconductor layers, generating a measurable electric current. Once the avalanche is triggered, the sensor needs to be quenched and restored to the original voltage so it can detect the next photon. The interval between quenching and restoration is the sensors hold-off time $\sDeadTime$, in which the diode cannot detect other incoming photons. Special phenomena must be taken into account due to external and internal noise or material flaws \cite{Renker2006,Zappa2007}. 

The presented model is defined as a probabilistic model, that simulates most processes occurring in a SPAD as photons arrive. As such, it is defined as a set of probability density functions (pdf), that are sampled in run time. Given the effect each photon on the following ones (due to e.g. SPAD's dead time after quenching), we cannot model the photon arrival as independent stochastic processes, but as a Markovian process. 
In the following, we describe the different effects of a SPAD, together with their associated pdfs. Later, we explain how these effects are combined into our probabilistic SPAD computational model.

\subsection{Photon detection efficiency}

In order to be detected, a photon must be absorbed and then trigger the avalanche process. The probability of triggering the avalanche is modelled by the photon-detection efficiency $E$, defined as the ratio between the number of incoming photons and the number of output pulses as 
\begin{equation}
E(\lambda,V_e)=\eta(\lambda)\cdot P_T(V_e),
\label{eq:PDE}
\end{equation}
where $\eta$ is the SPAD absorption efficiency, and $P_T(V_e)$ is the avalanche trigger probability \cite{Savuskan2013}. 
The absorption efficiency $\eta(\lambda)$ is related to the physical configuration of the sensor and its material properties, and varies depending on the wavelength $\lambda$ of the incident radiation. It is given by 
\begin{equation}
\eta(\lambda) = (1-R)\,\exp(\alpha(\lambda) D)\,(1-\exp(\alpha(\lambda) W))
\end{equation}
where $\alpha(\lambda)$ is the silicon absorption coefficient, $W$ the depletion region thickness, $D$ the junction depth and $R$ the power reflection coefficient for and air/silicon interface \cite{Zappa2007}.

On the other hand, the avalanche trigger probability $P_T(V_e)$ is a function of the excess voltage of the diode $V_e=V-V_{BD}$, and it is given by 
\begin{equation}
P_{T}(V_e) = P_e(V_e)+P_h(V_e)-P_e(V_e)P_h(V_e),
\end{equation}
where $P_e(V_e)$ and $P_h(V_e)$ are the probability that an electron and hole respectively will produce an avalanche at $x$~\cite{Mcintyre1973}. These probabilities are proportional to $V_e$, and can be calculated by using a set of differential equations (see \cite{Mcintyre1973} for the full formulation). However, for low excess voltage the avalanche probability $P_{T}$ can be approximated using a semi-empirical formula~\cite{Dautet1993,Ghioni1996}
\begin{equation}
P_T(V_e)\approx 1-\exp(-V_e/V_c),
\end{equation}
where $V_e$ is the excess voltage and $V_c$ is the characteristic voltage.

\begin{figure}[t]
  \centering
  \includegraphics[width=3.0in]{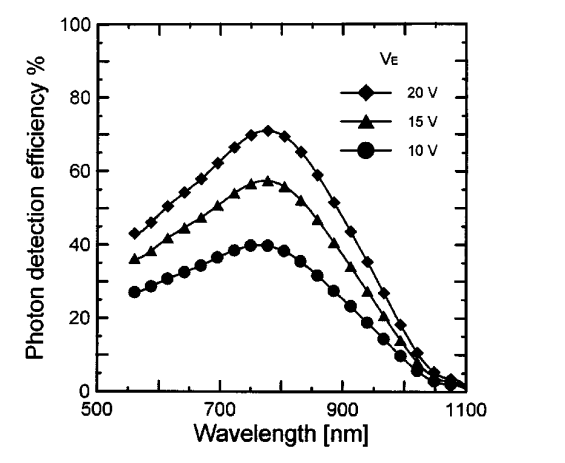}
  \caption{Photon detection efficiency $E(\lambda,V_e)$ as a function of wavelengths for different excess voltage bias. Image from \protect\cite{Cova1996}.}
  \label{fig:PDE}
\end{figure}

Figure \ref{fig:PDE} shows different photon detection efficiency curves as a function of wavelenght for different excess voltages. In our experiments we set both the excess voltage and wavelength, which allows us to use a fixed efficiency $E$.

\subsection{Time jitter}
As photons hit the sensor, their arrival time is captured within an error interval due to the electronics of the SPAD. This error is termed \emph{time jitter}, and is defined as the difference between the real photon arrival time and when it is recorded. 
This is a very important parameter as it defines the temporal maximum resolution of the sensor (i.e. in particular its point spread function). In general, time jitter $\timeJitter$ is modeled as a characteristic curve obtained with a time-correlated single-photon counting (TCSP) device~\cite{Oconnor2012}. This curve has two well-defined parts (see Figure~\ref{fig:jitter}): a Gaussian peak followed by an slower exponential tail as
\begin{equation}
\timeJitter\sim G(\mu,\sigma)+Exp(\tau).
\end{equation}
The Gaussian term $G$ is usually defined by the full-width at half-maximum (FWHM) of the curve $\FWHM$, which can be directly transformed to the standard deviation using
\begin{equation}
\sigma=\frac{\FWHM}{2\sqrt{2\ln(2)}},
\end{equation}
while the mean value $\mu$ is the time when most photons reach the sensor (the maximum count value of the curve). The time constant $\tau$ of the exponential distribution is computed following 
\begin{equation}
\tau=\frac{W_{n}^2}{\pi D_e},
\end{equation}
where $W_{n}$ is the thickness of quasi-neutral p region and $D_e$ is the diffusion coefficient of electrons \cite{Lacaita1989}.

Time jitter has been analytically implemented using an exponentially modified Gaussian distribution~\cite{Jeansonne1991} or by linear blending between both independent distributions~\cite{Zhang2009}. However, since we have access to actual measurements of the sensor, we choose for a data driven approach where $\timeJitter$ is implemented as a tabulated pdf.  In order to create a pdf, we filter out the noise, and tabulate and normalize the curve. 

\subsection{Quenching \& Hold-off Time}

Once an avalanche is triggered, then the SPAD cannot detect photons. To restore the diode to operating levels, the avalanche must be quenched by lowering the bias below breakdown voltage. This process can be done passively or actively \cite{Tisa2007}. Passive circuits are simpler to implement physically, they are prone to errors, imposing a lower temporal resolution and thus reducing their applicability. Here we focus on active quenching circuits. 

%

Active quenching relies on a sophisticated electronic circuit which forces quenching and resets transition in much shorter times. This is achieved by triggering a fast comparator and driving an inverse voltage to the diode in order to force the avalanche to extinct. This makes the system much faster and reliable as the hold-off time $\sDeadTime$ is constant in every avalanche process and can be easily adjusted. This hold-off time $\sDeadTime$ imposes the non-independence between photon arrival events, transforming our probabilistic model on a Markovian one where the result of each event is dependent on previous states in the chain.

\subsection{Internal noise} 
Due to their high-sensibility SPADs suffer from errors in form of sensor internal noise, that might trigger incorrect photon counts and therefore degrade the signal. This noise is visible as three effects mainly: dark counts, afterpulsing and crosstalk between neighbor pixels. 

\subsubsection{Dark counts}

\begin{figure}[t]
  \centering
  \includegraphics[width=3.0in]{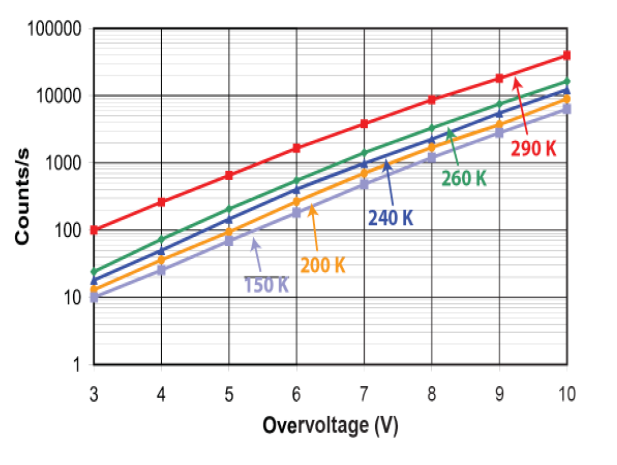}
  \caption{Exponential behavior of dark counts as a function of the excess bias voltage and the temperature. Figure from \protect\cite{Tisa2007b}.}
  \label{fig:DC}
\end{figure}

Dark counts are thermally generated carriers that can trigger an avalanche process even when the device is operating in dark conditions, resulting in a false count. The probability of dark counts $P_{dc}$ is directly related to the thermal (temperature) and electrical (excess bias voltage) energy of the system, following an exponential relationship~\cite{Tisa2007b}, as shown in Figure \ref{fig:DC}.

Dark counts are discrete events in a fixed interval of time, well described as a Poisson distribution with mean value the so-called dark count rate $\DCR$:
\begin{equation}
P_{dc}\sim \Poisson(k)=\frac{\DCR^k \exp(-\DCR)}{k!}.
\label{eq:dcr}
\end{equation}

\subsubsection{Afterpulsing}

\begin{figure}[ht]
  \centering
  \includegraphics[width=3.0in]{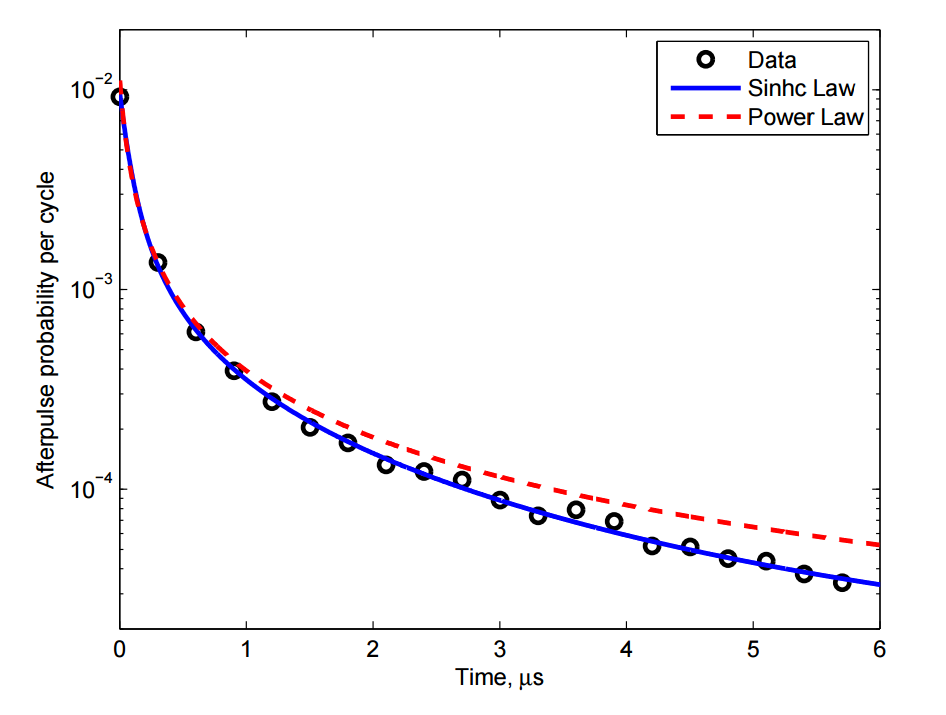}
  \caption{Afterpulse probability $P_{ap}$ as a function of the hold-off time $\sDeadTime$. Source: \protect\cite{Horoshko2014}.}
  \label{fig:APP}
\end{figure}

Another source of internal noise, which generates false photon counts, is afterpulsing. It is produced when previous carriers get trapped in the depletion layer, and are later released triggering an additional avalanche with a considerable delay. 
Afterpulse is modeled as the probability of an avalanche triggering another avalanche after the hold-off time ($\sDeadTime$).
The afterpulsing probability $P_{ap}$ is a function of time following an hyperbolic sinc function multiplied by a decreasing exponential function \cite{Horoshko2014} (see Figure~\ref{fig:APP}). Since in our implementation we fix the hold-off time $\sDeadTime$ we use a fixed afterpulse probability $P_{ap}$ for simplicity.

\subsubsection{Crosstalk}

\begin{figure}[t]
  \centering
  \includegraphics[width=3.0in]{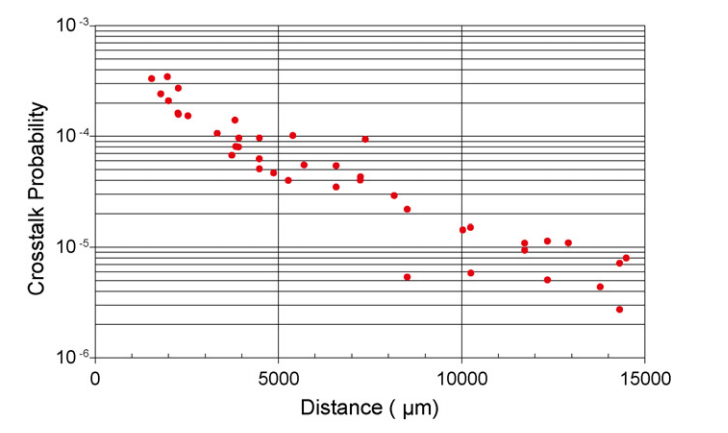}
  \caption{Measured crosstalk probability as a function of the pixel distance in $50\mu m$ SPAD sensors. Source: \protect\cite{Zappa2007}.}
  \label{fig:crosstalk2}
\end{figure}

In applications where 1D or 2D SPAD arrays are required, the electrical isolation of each pixel is crucial to avoid avalanche interference between adjacent pixels. This optical coupling or crosstalk can be significantly reduced with a correct pixel layout and highly doped isolations diffusions among pixels \cite{Zappa2007,Rech2008}.
%
%
However, even with a correct device design, a crosstalk probability $P_{ct}$ still exists. This crosstalk probability is dependent on the pixel's distance, as shown in Figure~\ref{fig:crosstalk2}. While crosstalk noise is rather low, we include it by probabilistic triggering an avalanche on neighbor pixels when a photon arrives to the sensor, based on $P_{ct}(d)$, with $d$ the distance between pixels. 

\subsection{External noise}
\begin{figure}[t]
  \centering
  \includegraphics[width=3.0in]{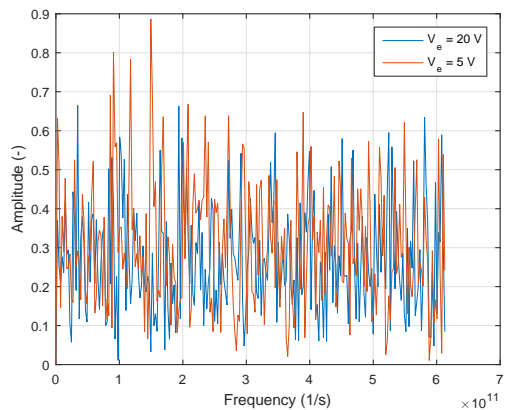}
  \caption{Fourier spectrum of the noise signals of a 20 $\mu m$ CMOS SPAD with excess voltages of 5 V and 20 V.}
  \label{fig:noise}
\end{figure}

Finally, in addition to the internal noise, SPADs are also sensitive to external noise due to ambient light and tunneling effects. This noise is visible in the tail at Figure~\ref{fig:jitter}, and it is independent on the SPAD's excess voltage. We analyze the frequency spectrum of the noise in two different noise signals from a 20 $\mu m$ CMOS SPAD sensor (with voltages at 5 V and 20 V, respectively), and found no dominant frequency (Figure~\ref{fig:noise}). Given the white-noise spectrum and discrete characteristics of the counting process, we model this noise as a Poissonian noise. In our experiments, we set the noise distribution mean $P_{e}$ to $5\cdot 10^{11}$ counts/s, calculated from the experimental data in Figure~\ref{fig:jitter}. 

\subsection{Implementation}

We implement our probabilistic model as a Matlab function taking as input a list of photons time of arrival, as well as their position in the case of multiple-pixel sensors. Algorithm~\ref{algtm:code} describes our model. For computational reasons, both the dark counts and the background noise are added at the end of the simulation, assuming that the probability of triggering afterpulse or crosstalk effects, as well as the effect of the hold-off time of noisy counts, are neglectable. 

\begin{algorithm}[h]
 \KwData{Ideal input photon arrival time ($t_{in}$), Simulation time ($t_{end}$)}
 \KwResult{Histogram of SPAD counts}
 initialization\;
 \For{all measurements}{
 initialize afterpulse time ($t_{ap}$)\;
  \For{all input photons}{      
  	\eIf{detected ($E$) {\bf and} not in hold-off time ($t_{in} > t_{ap}$) }{
   	$t$ = $t_{in}$ + $t_j$\;
   	sum $t$ to histogram\;
   	$t_{ap}$ = $t$ + $t_o$\;
   		\While{$t_{ap}$ $<$ $t_{end}$}{ 
   		\If {afterpulse ($P_{ap}$)}{
   		sum $t_{ap}$ to histogram\;
   		$t_{ap}$ = $t_{ap}$ + $t_o$\;}}
   		
   		\If{crosstalk ($P_{ct})$}{
   		sum crosstalk to histogram\;}
   		
   		go back to next input photon\;
   	}
   	{go back to next input photon\;}
  }
 }

 sum dark counts ($P_{dc}$) to histogram\;
 sum background noise ($P_{e}$) to histogram\;
 \
 \caption{Pseudo-code of our probabilistic SPAD model, transforming the ideal time of arrival of photons into the sensor's response.  }
 \label{algtm:code}
\end{algorithm}

%
%
%

\section{Results}


We evaluate our model under three different scenarios: a time-correlated single-photon counting (TCSP) process, a multiple impulse response, and a more complex temporal response including diffuse interreflections from simulations~\cite{Jarabo2014}.

\subsection{TCSP simulation}

\begin{figure}[ht]
  \centering
  \includegraphics[width=3.0in]{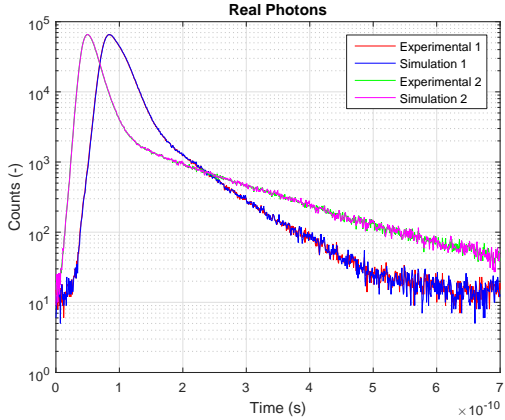}
  \caption{Time-correlated single-photon counting (TCSP) simulations of two different conditions in a SPAD detectors, one with narrow FWHM and other with a longer exponential tail. Our model accurately matches both the time jitter temporal PSF, as well as the signal noise.}
  \label{fig:sim1}
\end{figure}

Figure~\ref{fig:sim1} compares our model with respect to captured data for an impulse TCSP process. The captured data comes from measurements of a 20$\mu m$ diameter CMOS SPAD sensor with 7 V excess voltage. Two different device bias conditions are considered: a first one optimized for low jitter, i.e. narrow FWHM, and a second one optimized for fast tail. The simulation parameters are shown in Table \ref{tab:param}. Our method is able to accurately matching both the temporal response and the internal and external noise of the sensor. 

\begin{table}[ht]
\centering
\begin{tabular}{|c|c|}
\hline
Parameter & Value \\
\hline
\hline
Dark count rate & 3000 counts/s \\
\hline
Photon detection probability & 30 \% \\
\hline
Afterpulse probability & 1 \% \\
\hline
Dead time & 10 ns\\
\hline
Background noise & 14.7 counts/ps\\
\hline
Time jitter FWHM & 26 ps - 36 ps \\
\hline
Time jitter tail & 156 ps - 75 ps\\
\hline
Measurements & $\text{3.57}\cdot \text{10}^\text{6}$\\
\hline
\end{tabular}
\caption{Parameters of our SPAD model used in our simulations.}
\label{tab:param}
\end{table}

\subsection{Multiple impulse simulation}

\begin{figure}[ht]
  \centering
  \includegraphics[width=3.0in]{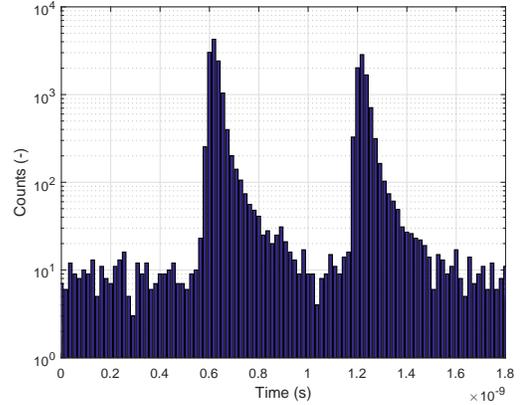}
  \caption{SPAD sensor response for two-impulse incoming signal, with both impulse responses have similar photon counts. Note that the first peak has slightly higher magnitude due to the hold-off time which in a perfect sensor sensitivity would completely mask the second peak of the signal. }
  \label{fig:sim2}
\end{figure}

We evaluate our method with two impulse responses arriving the sensor at times 0.6 and 1.2 ns. Given that the dead time of the SPAD is larger than the time resolution, several measurements must be done in order to capture the arrival of several photons with enough precision. Figure \ref{fig:sim2} shows the SPAD simulation as a histogram of $5\cdot 10^4$ measurements. 

As expected, the first impulse response has slightly more detections than the second one due to the photon detection probability and the hold-off time (with perfect photon detection the second impulse would not be captured). The dark count rate is neglectable at this time scales, as opposed to the background noise. Finally, it can be seen the characteristic Gaussian peak and exponential tail of the sensor's temporal PSF due to the time jitter. 

\subsection{Full transport simulation}

We additionally evaluate our SPAD model against a more complex light transport response, featuring direct illumination, as well as multiple diffuse interreflections. We use data obtained from time-resolved light transport simulation~\cite{Jarabo2014}. To simulate the sensor response we use the spatio-temporal response output by the transient renderer as a pdf of the incoming photons. 
\begin{figure}[ht]
  \centering
  \includegraphics[width=2.0in]{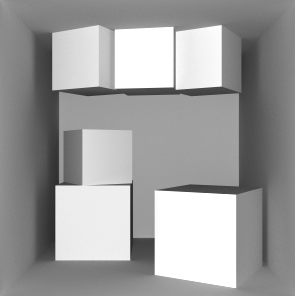}
  \caption{Steady-state render of the scene used to test our model, featuring both impulse response due to direct reflection, as well as multiple diffuse interreflections.}
  \label{fig:scene}
\end{figure}

\begin{figure}[ht]
  \centering
  \includegraphics[width=3.5in]{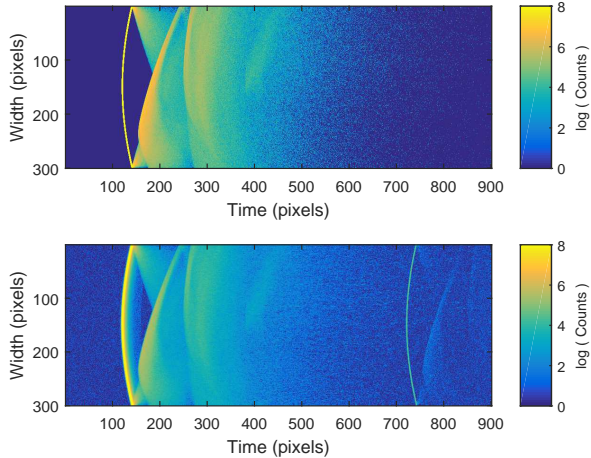}
  \caption{Rendered input spatio-temporal ideal response of the image (top), and the response of the SPAD for that signal. The x-axis represents time (where each pixel represents 16.6 ps), and the y-axis represents an horizontal scanline from Figure~\ref{fig:scene}. Color codes photon counts in logarithmic scale.}
  \label{fig:streak_spad}
\end{figure}

\begin{figure}[ht]
  \centering
  \includegraphics[width=3.0in]{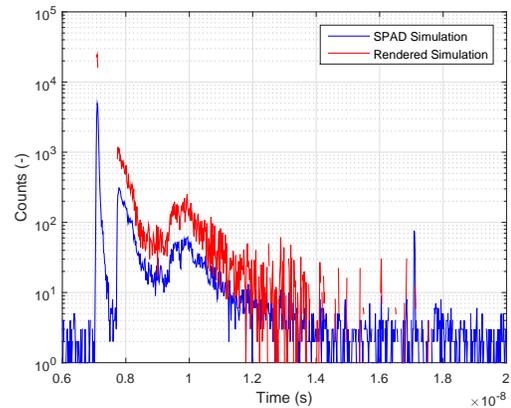}
  \caption{Single-pixel temporal response of the ideal simulated response of the scene at Figure~\ref{fig:scene} (red), and the response of the SPAD to that input image (blue). }
  \label{fig:scene_spad}
\end{figure}

Figure~\ref{fig:scene} shows an steady-state render of the scene used, while Figure~\ref{fig:streak_spad} (top) shows an example spatio-temporal response of light transport for a single scanline of the simulation, where the temporal resolution is set to 16.6 ps. 
%
%
Figure \ref{fig:streak_spad} (bottom) shows the SPAD response to the ideal incoming radiance on the scanline shown on top, with $10^4$ measurements per pixel. 
As expected, the time jitter of the sensor blurs the temporal response following the PSF defined by the time-jitter curve. This is clearly shown in the first wave front of the streak image. The afterpulsing is also clearly seen as a repetition pattern of the main wavefronts arrival times. It has to be taken into account that the afterpulse probability was set to $P_{ap}=1\%$, meaning that only $1\%$ of the photons could potentially trigger a false avalanche. The probability of a $n^{th}$ afterpulse decreases exponentially with probability $P_{ap}^n$. Note also that the intensity of the diffuse reflections is reduced with respect to the original signal due to hold-off time. 
Finally, Figure~\ref{fig:scene_spad} plots the temporal response of a single pixel to directly compare between the incoming ideal signal and the sensor's response. It shows how the incoming signal exponentially decreases along time due to the sensor's detection probability and hold-off time, and the added background noise. Note also the afterpulse depicted as a small peak at instant 1.7 ns. 

\section{Conclusions}
In this work we have presented a computational model for single-photon avalanche diodes (SPAD). We posed our model as a probabilistic Markovian model, modeling the sensor response to a set of non-independent photon arrivals. Our work includes most of the effects of SPADs on the output response, including the detection efficiency, time jitter, avalanche quenching, as well as the sensor's internal and external sources of noise. The goal of our work is to provide an accurate computational sensor model, to be used on top of physically-based transient light transport simulations: This is important to use these simulations on the development of new computational techniques for image understanding on transient imaging, where SPAD sensors have emerged as a promising low-cost and highly-efficient imaging technology. 

\section*{Acknowledgements}
We want to thank Alberto Tosi and his team at Politecnico di Milano for providing references on SPADs theory and characterization, as well as measured data from real-world SPADs. 
This research has been funded by DARPA (project REVEAL), and the Spanish Ministerio de Econom\'{i}a y Competitividad (projects TIN2016-78753-P and TIN2014-61696-EXP).

\bibliographystyle{acmsiggraph}
\nocite{*}
\bibliography{ms}
\end{document}